\documentclass{iaus}
\usepackage{graphicx}

\title[Wind-shearing of planetesimals]
{Wind-shearing in gaseous protoplanetary disks}

\author[Perets \& Murray-Clay]   
{Hagai B. Perets \& Ruth Murray-Clay}

\affiliation{Harvard-Smithsonian Center for Astrophysics, 60 Garden st. Cambridge MA 02338, US}

\pubyear{2010}
\volume{276}  
\pagerange{???}
\jname{Wind-shearing of planetesimals in gaseous protoplanetary disks}
\editors{}
\begin{document}

\maketitle

\begin{abstract}
One of the first stages of planet formation is the growth of small
planetesimals and their accumulation into large planetesimals and
planetary embryos. This early stage occurs much before the dispersal
of most of the gas from the protoplanetary disk. Due to their different
aerodynamic properties, planetesimals of different sizes/shapes experience
different drag forces from the gas at these stage. Such differential
forces produce a wind-shearing effect between close by, different
size planetesimals. For any two planetesimals, a wind-shearing radius
can be considered, at which the differential acceleration due to the
wind becomes greater than the mutual gravitational pull between the
planetesimals. We find that the wind-shearing radius could be much
smaller than the \emph{gravitational} shearing radius by the Sun (the
Hill radius), i.e. during the gas-phase of the disk wind-shearing
could play a more important role than tidal perturbations by the Sun.
Here we study the wind-shearing radii for planetesimal pairs of different
sizes and compare it with gravitational shearing (drag force vs. gravitational
tidal forces). We then discuss the role of wind-shearing for the stability
and survival of binary planetesimals, and provide stability criteria
for binary planetesimals embedded in a gaseous disk. 
\end{abstract}

The interactions between planetesimals play an important role in the
evolution of protoplanetary disks and planet formation (Lissauer 1993, Goldreich et al. 2004).
Most of these interactions and the growth of planetesimals likely
occur while the planetesimals are embedded in a gaseous disk. Here we
focus on the close interaction between pairs of single planetesimals,
i.e. planetesimal-planetesimal-gas interactions. 

Planetesimals could vary in size and shape, and therefore have a wide
range of aerodynamical properties, which would affect their interaction
with the surrounding gas. In particular, planetesimals of different
sizes/shapes experience different drag forces from the head wind they
encounter in the gaseous disk. The difference between the forces acting
on two, different size planetesimals, could effectively change their
relative trajectories in respect to their unperturbed motion in the
absence of gas (see also Ormel \& Klahr 2010). In particular, during
an encounter between two different size planetesimals such differential
forces result in a wind-shearing (WISH) effect, which could be even
stronger than their gravitational interaction. For any two planetesimals,
a WISH radius can be considered, at which the acceleration due to
aerodynamical wind-shearing becomes greater than the mutual gravitational
pull between them. Binary planetesimals can not survive with separation larger than the WISH radius, even if they are gravitationally bound. These issues are discussed in detail in Perets \& Murray-Clay (in preparation).

The differential acceleration between
two planetesimal of radii $r_b$ and $r_s$, due to the wind-shearing
effect is given by $\Delta a=|F_D(r_b)/m_b-F_D(r_s)/m_s|$,
 where $F_{D}(r)$ is the drag force affecting a planetesimal of radius $r$. 

For small distances, at which the environmental conditions 
are approximately the same, the differential WISH  between
any two planetesimals is independent of the distance between them.
However, similar to the the Hill radius which arises from the tidal
graviataional shearing from the Sun, we can define an important relevant
distance scale; the wind-shearing radius. We define this radius as the distance
between two planetesimals for which the differential WISH acceleration
between them equals their mutual gravitational pull. Beyond this limiting
radius even two planetesimals which are gravitationally bound
(in the absence of WISH) would be sheared apart by the wind.  

We can now derive the WISH radius for any given pair of planetesimals,
 by equating $\Delta a_{WS}$
with the gravitational acceleration $a_{grav}=GM/r_{WS}^{2}\label{eq:a_grav}$ 
at the WISH radius ($r_{WS}$). 

\begin{figure}
\includegraphics[scale=0.56]{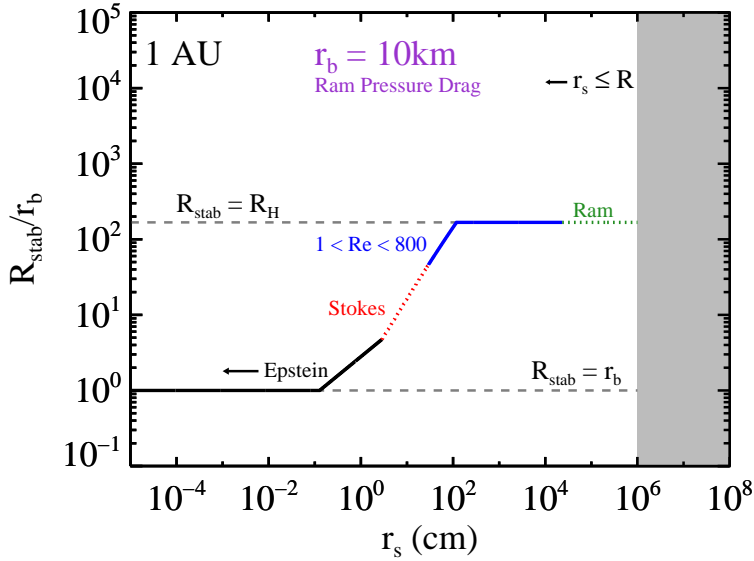}\caption{\label{fig:WISH-stability}The WISH radius of a $10$
km planetesimal and a planetesimal of size $r_{s}$, as a function
of $r_{s}$ at 1 AU from the star.  The physical size and the Hill radii of
the large size planetesimal (lower and upper dashed lines, respectively),
are also shown. Also shown are the appropriate drag regimes, which depend on the Reynolds number Re.}

\end{figure}

The WISH radius can be found for any two planetesimals of arbitrary
size, as illustrated in fig. \ref{fig:WISH-stability}. 
This figure shows the calculated WISH
radius as a function of the small planetesimal size. The transition
between different drag regimes can be seen in this figure. 

Binary planetesimals could be strongly affected by gas drag. In a
gas free environment, binary planetesimals are stable as long as their
separation is smaller than the Hill radius, whereas wider binaries
are destabilized and disrupted by the tidal gravitational shearing
from the Sun. However, when gas drag is taken into account, the Hill
radius stability limit should be replaced by the WISH radius (as long
as $r_{WS}<r_{Hill};$ binaries wider than the Hill radius are always
unstable), i.e. we can formulate a stability criteria for binaries
embedded in gas: $a_{bin}\le min(r_{Hill},r_{WS})$.
In Fig. \ref{fig:WISH-stability} the WISH radius
represents the limiting separations of binary planetesimals ($R_{stab}$) or small satellites embedded in a gaseous environment at ahich bound binary planetesimals can exist. The appropariate phase space is delimited by the physical radii, the WISH radii, and the Hill radii. 
 
\bibliographystyle{plainnat}

\end{document}